\documentclass[preprint,12pt]{elsarticle}
\usepackage{latexsym, graphicx, epsfig, amsmath, amssymb, amsfonts}
\usepackage{amsmath}
\usepackage{amssymb}
\usepackage{graphicx}
\usepackage{subfig}
\usepackage{tabu}
\usepackage{adjustbox}
\usepackage{bbold}
\usepackage{bm}
\usepackage{textcomp}
\usepackage[thinc]{esdiff}
\usepackage[dvipsnames]{xcolor}
\usepackage{comment}
\usepackage{textgreek}
\usepackage{caption}
\usepackage{lineno}
\usepackage{multirow}
\usepackage{adjustbox}

\usepackage{siunitx}
\usepackage{xcolor}
\usepackage{booktabs}
\usepackage{caption}
\usepackage{float}
\usepackage{color,soul}

\begin{document}
\begin{frontmatter}

\title{Domain Structure and Interface Control of Mechanical Stiffness in Sustainable Cellulose Bio-nanocomposites}

\author{Hanxun Jin\textsuperscript{a,b,c,d}}
\author{William Goldberg\textsuperscript{a,b,d}}
\author{Zhenqin Wang\textsuperscript{a,b,c}}
\author{Huiyong Li\textsuperscript{a,b,c}}
\author{Yuxuan Huang\textsuperscript{a,b,d}}
\author{Marcus Foston\textsuperscript{a,b,c,}\corref{cor1}}
\author{Guy M. Genin\textsuperscript{a,b,d,}\corref{cor2}}

\cortext[cor1]{Corresponding author. E-mail:  mfoston@wustl.edu (M.F.)}
\cortext[cor2]{Corresponding author. E-mail:  genin@wustl.edu (G.M.G.)}

\address[1]{SMARC, The Synthetic-Biology Manufacturing of Advanced Materials Research Center, Washington University in St. Louis, Saint Louis, Missouri 63130, United States}
\address[2]{NSF Science and Technology Center for Engineering MechanoBiology, Washington University in St. Louis, Saint Louis, Missouri 63130, United States}
\address[3]{Department of Energy, Environmental and Chemical Engineering, Washington University in St. Louis, Saint Louis, Missouri 63130, United States}
\address[4 ]{Department of Mechanical Engineering \& Materials Science, Washington University in St. Louis, Saint Louis, Missouri 63130, United States}

\begin{abstract}
Renewable and biodegradable plastics derived from soy protein isolate (SPI) offer a promising alternative to conventional petroleum-based plastics, particularly for film-grade bioplastics applications such as plastic bags. However, even with reinforcement from cellulose nanocrystals (CNCs), their mechanical properties including stiffness lag behind those of petroleum-based plastics. To identify pathways for improving CNC-reinforced SPI composites, we studied stiffening mechanisms by interpreting experimental data using homogenization models that accounted for CNC agglomeration and the formation of CNC/SPI interphases. To model  effects of surface modification of CNCs with polydopamine (polyDOPA), we incorporated two key mechanisms: enhanced CNC dispersion and modified CNC-SPI interfacial interactions. Models accounted for interphases surrounding CNCs, arising from physicochemical interactions with the polyDOPA-modified CNC surfaces. Consistent wih experimental observations of polyDOPA modification enhancing mechanical properties through both increased spatial distribution of CNCs and matrix-filler interactions, results demonstrated that improved dispersion and interfacial bonding contribute to increased composite stiffness. Results highlight the potential of biodegradable CNC/SPI bio-nanocomposites as sustainable plastic alternatives, and suggest pathways for further enhancing their mechanical properties.
\end{abstract}


\end{frontmatter}

\section{Introduction}
The global dependence on petroleum-based polymers has created pressing environmental challenges due to their persistence in ecosystems and slow degradation rates \cite{gowthaman2021, jin2022, epps2021,jin2021ruga, zhao2023}. 
This issue has sparked growing interest in developing sustainable alternatives, particularly biodegradable polymers reinforced with natural materials \cite{ncube2020, mukherjee2023}. Such materials address increasing global demand for environmental sustainability while offering potential solutions for certain low-strength engineering applications \cite{mohanty2022, samir2022, calvino2020}. Among promising sustainable options, soy protein isolate (SPI), derived from soybeans, can be polymerized into film-grade bioplastics \cite{song2011}.
However, SPI’s inherent brittleness limits its usability in many plastics applications. Adding plasticizers like glycerol improves ductility but simultaneously compromises tensile strength, highlighting the need for reinforcement strategies \cite{trache2017}.

\begin{figure}[h]
    \centering
    \includegraphics[width=1.0\textwidth]{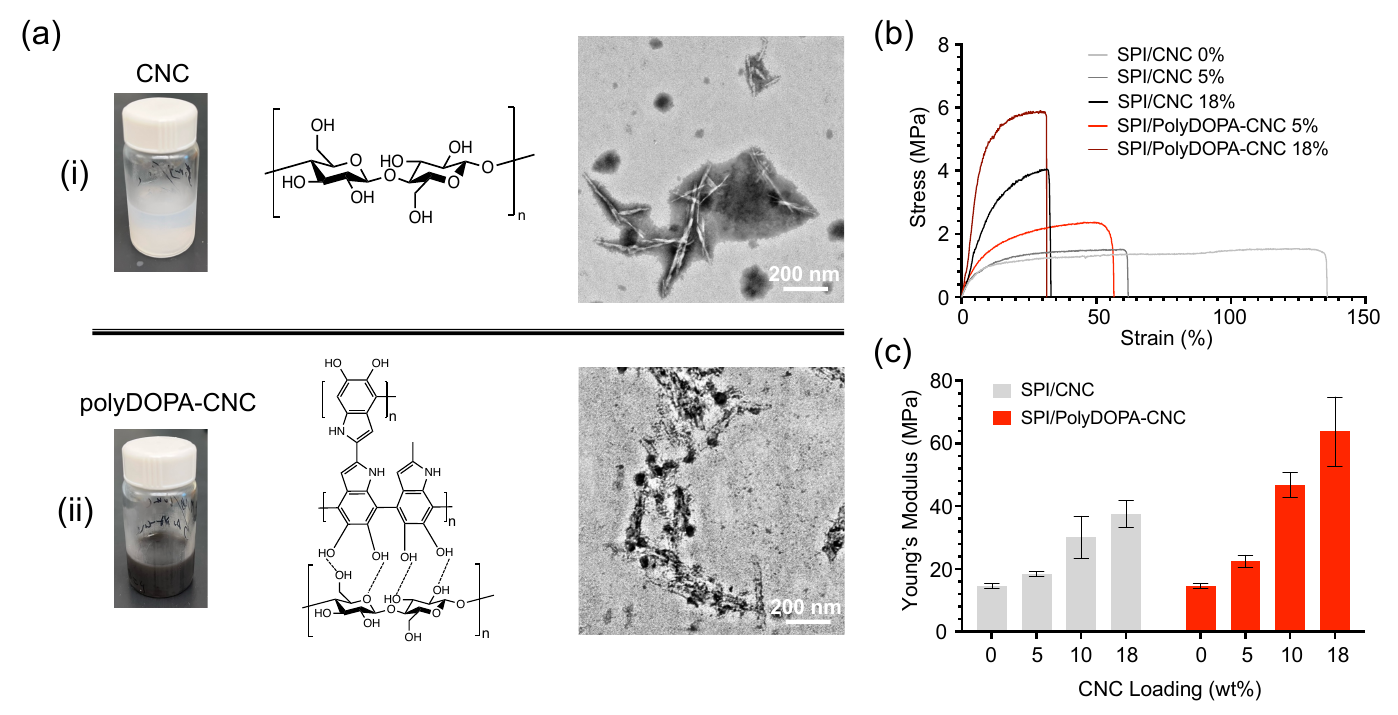}
\caption{\textbf{Mechanical characterization of cellulose nanocrystal (CNC) reinforced soy protein isolate (SPI)-glycerol composites.} 
(a) Chemical structure and transmission electron microscope (TEM) images showing how CNCs change with surface modification: (i) unmodified CNC regions appear as long, needle-like structures, while (ii) after coating with polydopamine (polyDOPA), CNC regions appear shorter and thicker. Dotted lines represent hydrogen bonds between molecules. 
(b) Stress-strain measurements showing how the material stretches and deforms. Each curve represents a different concentration (weight percentage, Wt\%) of CNCs. Higher CNC content makes the material stiffer but reduces the failure strain. 
(c) Summary of material stiffness (Young's modulus) for different CNC concentrations, comparing unmodified CNCs (SPI/CNC) to surface-modified CNCs (SPI/PolyDOPA-CNC). Surface modification with polyDOPA significantly increases material stiffness. Figure panels reproduced with permission from \cite{wang2024}.}
    \label{fig:1}
\end{figure}

Cellulose nanocrystals (CNCs), produced through acid hydrolysis of plant-based agricultural residues such as soy leaves and stalks, offer a sustainable solution for enhancing SPI's mechanical properties, as they are strong, naturally abundant, and biodegradable \cite{shojaeiarani2019, rashid2023, cao2021cellulose, habibi2010}. CNCs combine excellent mechanical properties with nanoscale dimensions: they are slender (2–20 nm in diameter and hundreds of nanometers to micrometers in length) \cite{moon2011} with an axial elastic modulus of 50-200 GPa that rivals steel \cite{channab2024} (\textbf{Fig.\ref{fig:1}a(i)}). When well-dispersed, these nanofibers can effectively stiffen polymer matrices by promoting local chain confinement and enhancing crystallization \cite{zhang2008polymer, zhu2020spatiotemporally}.

Despite their potential, challenges remain in optimizing CNC-based nanocomposites. A fundamental challenge lies in the competing effects of CNC surface properties. While high surface energy promotes beneficial matrix interactions and stress transfer \cite{gomri2022, habibi2014, abraham2016}, it simultaneously drives CNC aggregation that can degrade mechanical performance \cite{chu2020}. Surface modification strategies, such as polydopamine (polyDOPA) coating \cite{zhang2016, rocha2023, xu2023} (\textbf{Fig.\ref{fig:1}a(ii)}), aim to enhance matrix compatibility but often result in either CNC degradation or increased aggregation \cite{peng2017, babaei-ghazvini2023}.

PolyDOPA-modified CNCs increase the stiffness of SPI-glycerol nanocomposites but simultaneously reduce failure strain (\textbf{Fig.\ref{fig:1}b}) \cite{wang2024}. This polyDOPA surface modification of CNCs can increase composite stiffness by more than 1.5 times compared to unmodified CNCs at equivalent weight fractions (\textbf{Fig.\ref{fig:1}c}), possibly due to modifications of microstructural properties and CNC-matrix interactions that are governed by non-bonded electrostatic and van der Waals forces \cite{miao2019,wang2024}. Transmission electron microscopy (TEM) analysis further confirmed significant CNC aggregation, which can reduce mechanical stiffness \cite{wang2024}.

The interplay between beneficial CNC-matrix interactions and detrimental CNC aggregation highlight the need for a deeper understanding of the nanoscale mechanics governing CNC-reinforced bio-nanocomposites. Previous investigations have employed classical micromechanics approaches, such as the Halpin-Tsai \cite{capadona2008, favier1995} and Mori-Tanaka models \cite{demir2022, josefsson2014}, to analyze specific CNC-based composites. These first-order approaches often treat CNCs as perfectly dispersed, high-aspect-ratio inclusions within a homogeneous matrix, highlighting the limitations of these approaches in capturing the complexities of CNC aggregation and the heterogeneous interphase properties. The primary challenge in the modeling lies in simultaneously capturing multiple competing mechanisms: CNC aggregation, interphase effects, and the influence of surface modifications. Additionally, the significant modulus mismatch between CNCs and typical biopolymer matrices (often exceeding 10,000:1) \cite{weng1984, qiu1990} necessitates careful consideration of load transfer efficiency. To address these challenges, we developed a micromechanics model that explicitly accounts for both domain agglomeration and interfacial interactions. This framework provides broader understanding and offers guidelines for designing sustainable CNC-reinforced bio-nanocomposites with enhanced mechanical properties.

\section{Modeling}

The development of predictive models for CNC-reinforced bio-nanocomposites requires addressing three key challenges identified through study of their experimental characterization \cite{wang2024}: (1) significant CNC aggregation even at low weight fractions, (2) complex interactions at CNC-matrix interfaces that are modified by surface treatments, and (3) a substantial modulus mismatch between CNCs and the SPI-glycerol matrix that affects load transfer. To address these challenges, we first developed a model for ideally dispersed CNCs to establish theoretical bounds, then incorporated effects of aggregation, and finally accounting for interphase regions reported in experimental observations of polyDOPA-modified CNCs. 

\subsection{Stiffness of Nanocomposites with Fully Dispersed CNCs}

\begin{figure}[h]
    \centering
    \includegraphics[width=0.6\textwidth]{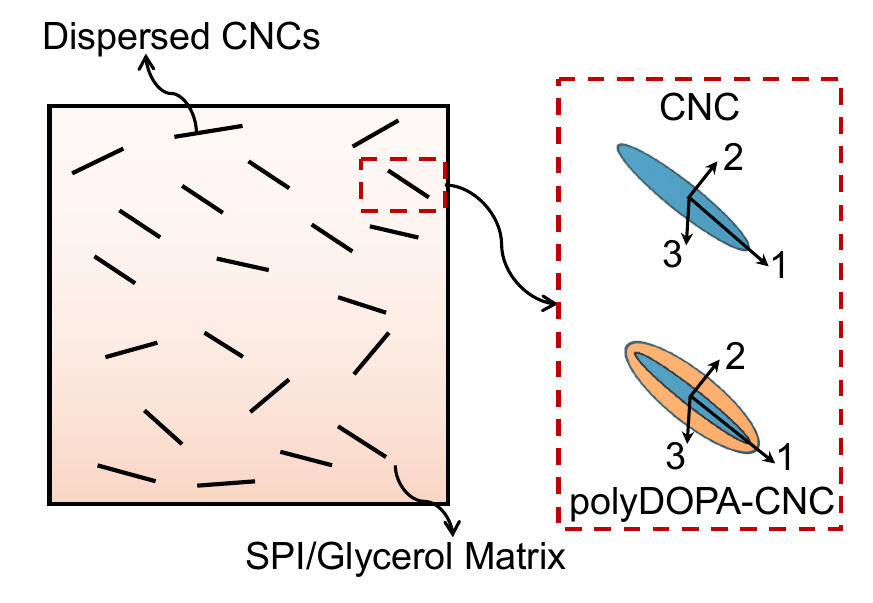}
    \caption{\textbf{Schematic of an idealized model for CNC reinforcement in a SPI-glycerol matrix.} The model for the idealized case assumes perfectly dispersed CNCs (unmodified or polyDOPA-modified) within a continuous matrix phase. This idealized case serves as a theoretical baseline for understanding the effects of CNC surface modification and evaluating the impact of agglomeration and interface quality on composite mechanics. PolyDOPA-modified CNCs are shown with their characteristic surface coating, which affects both interfacial interactions and dispersion behavior.}
    \label{fig:2}
\end{figure}

Although TEM analysis of SPI/CNC composites reveals significant CNC agglomeration even at 5 Wt\% CNC loading \cite{wang2024}, surface treatment enhances dispersion within the matrix \cite{capadona2008}. We first considered an idealized scenario where high-aspect-ratio CNCs are uniformly dispersed within the matrix (\textbf{Fig.~\ref{fig:2}}). As appropriate for intermediate, sub-percolation CNC volume fractions \cite{mori1973,saadat2015effective,genin2009micromechanics}, We followed the Mori-Tanaka approach, where the representative volume element (RVE) consists of a continuum with CNC inclusions of volume fraction $V_c$ perfectly bonded to a homogeneous SPI-glycerol matrix of volume fraction $V_m = 1-V_c$. In this method, each inclusion experiences the matrix strain as its far-field strain rather than the overall composite strain. The ensemble average of the strain in the matrix phase acts as the reference strain, and the strain inside each inclusion is related to the matrix strain through the dilute strain concentration tensor. The effective properties are then determined by volume averaging the fields over all phases, considering their interactions through the matrix mean-field.
The Mori-Tanaka effective elastic stiffness tensor of the nanocomposite, \( \bm{\overline{C}} \), is:
\begin{equation}
    \bm{\overline{C}} = \bm{C}_m + V_c (\bm{C}_c - \bm{C}_m) : \bm{A}_c.
\end{equation}
where bold symbols represent second or fourth-order tensors; the operator ``:'' indicates an inner product; \( \bm{C}_m \) and \( \bm{C}_c \) are the elastic moduli of the matrix and CNCs, respectively; and the concentration tensor \( \bm{A}_c \) relates \( \bm{\varepsilon}_0 \) to the average strain in CNCs \( \bm{\varepsilon}_c \), as \( \bm{\varepsilon}_c = \bm{A}_c \bm{\varepsilon}_0 \).
\( \bm{A}_c \) is expressed as:
\begin{equation}
    \bm{A}_c = \left[ \mathbf{I} - \mathbf{S} : \mathbf{C_m}^{-1} : (\mathbf{C_m} - \mathbf{C_c}) \right]^{-1}.
\end{equation}
where \( \bm{I} \) is the identity tensor, and \( \bm{S} \) is the Eshelby tensor \cite{eshelby1957}, which depends on the inclusion’s shape and orientation. Therefore, \( \bm{\overline{C}}\) can be estimated from the stiffness tensors of the two phases (\( \bm{C}_m, \bm{C}_c \)) and \( \bm{S} \) for the specific inclusion geometry and orientation.

Although transverse properties of even highly anisotropic long, slender fibers do not typically contribute strongly to the mechanics of composites \cite{genin1997composite,maurin2008transverse}, we explored the possibility that transverse isotropy of CNCs or CNC bundles might influence the elastic modulus; this effect proved negligible for SPI/CNC nanocomposites as well, even when the transverse elastic modulus differed by an order of magnitude (\textbf{Fig.~\ref{fig:S1}}).
For this overly general case, with the 1-direction aligned along the axis of the CNC, the elastic stiffness matrix \( \bm{C}_c \) can be expressed using Hill’s notation \cite{hill1964a, hill1964b, hill1965} as:

\begin{equation}
\bm{C}_c = 
\begin{bmatrix}
n_c & l_c & l_c & 0 & 0 & 0 \\
l_c & k_c + m_c & k_c - m_c & 0 & 0 & 0 \\
l_c & k_c - m_c & k_c + m_c & 0 & 0 & 0 \\
0 & 0 & 0 & p_c & 0 & 0 \\
0 & 0 & 0 & 0 & p_c & 0 \\
0 & 0 & 0 & 0 & 0 & m_c
\end{bmatrix}.
\end{equation}
where $n_c = {E_c^{11} (1 - \nu_{23})}/({1 - 2\nu_{12} \nu_{21} - \nu_{23}})$, $k_c = {E_c^{11}}/[{2(1 - 2\nu_{12} \nu_{21} - \nu_{23})}],
l_c = 2\nu_{12} k_c$, $
p_c = {E_c^{11}}/{[2(1 + \nu_{12})]}$,  and 
$m_c = {E_c^{22}}/{[2(1 + \nu_{23})]}$. 
For unmodified CNCs, \( E_c^{11} = 50 \text{ – } 200 \, \text{GPa} \), and \( E_c^{22} = E_c^{33} = 0.1E_c^{11} \) \cite{moon2011}. We set \( \nu_{12} = \nu_{23} = \nu_{13} = 0.25 \), well below the thermodynamic limits.

For CNCs coated with a layer of polyDOPA (10-15 nm thickness), we used the Voigt model to estimate the effective moduli, $E_{\text{polyCNC}}^{ii} = E_{\text{poly}}^{ii} V_{\text{poly}} + E_c^{ii} (1 - V_{\text{poly}})$. Here, \( i = 1 \) or \( 2 \). \( E^{ii}_{\text{poly}} = 2.5 \, \text{GPa} \) is the isotropic elastic modulus of the polyDOPA layer \cite{li2019}. \( V_{\text{poly}} = 0.8 \) is the volume fraction of the polyDOPA coating, which was estimated from published TEM images \cite{wang2024}.

For randomly dispersed and oriented CNCs, the bulk modulus \( K \) and shear modulus \( G \) of the nanocomposite was writen following the approach of Benveniste \cite{benveniste1987},
\begin{equation}
    K = K_m + \frac{V_c (\delta_c - 3K_m \alpha_c)}{3(V_m + V_c \alpha_c)},
\end{equation}
\begin{equation}
    G = G_m + \frac{ V_c(\eta_c - 2G_m \beta_c)}{2(V_m + V_c \beta_c)}.
\end{equation}
where \( K_m \) and \( G_m \) are the bulk modulus and shear modulus of the SPI-glycerol matrix, and \( \alpha_c \), \( \beta_c \), \( \delta_c \), and \( \eta_c \) are \cite{shi2004},

\begin{equation}
    \alpha_c = \frac{3(K_m + G_m) + k_c - l_c}{3(G_m + k_c)},
\end{equation}
\begin{equation}
\beta_c = \frac{1}{5} \left[ \frac{4G_m + 2k_c + l_c}{3(G_m + k_c)} + \frac{4G_m}{G_m + p_c} + \frac{2G_m (3K_m + G_m) + 2G_m (3K_m + 7G_m)}{G_m (3K_m + G_m) + m_c (3K_m + 7G_m)}\right],
\end{equation}
\begin{equation}
\delta_c = \frac{1}{3} \left[ n_c + 2l_c + \frac{(2k_c + l_c)(3K_m + 2G_m - l_c)}{G_m + k_c} \right],
\end{equation}
\begin{equation}
\eta_c = \frac{1}{5} \left[ \frac{2}{3}(n_c - l_c) + \frac{8G_m p_c}{G_m + p_c} + \frac{8m_c G_m (3K_m + 4G_m)}{3K_m (m_c + G_m) + G_m (7m_c + G_m)} + \frac{2(k_c - l_c)(2G_m + l_c)}{3(G_m + k_c)} \right].
\end{equation}

Finally, the Young’s modulus \( E \) of the nanocomposite was calculated as,
\begin{equation}
    E = \frac{9KG}{3K + G}.
\end{equation}

Although the SPI-glycerol matrix can form biphasic nanostructures with intertwining SPI-rich and glycerol-rich phases ranging in size from 1 to 5 nm, as observed in previous TEM analysis \cite{tian2018}, the diameters of the CNCs, which range from 10 to 20 nm for unmodified CNCs and 30 to 50 nm for polyDOPA-CNCs, are significantly larger than the intertwining phase size. Therefore, we treated the SPI-glycerol matrix as isotropic, with Young’s modulus \( E_m = 14.7 \, \text{MPa} \) and the Poisson ratio \( \nu_m = 0.45 \) \cite{wang2024}.

\subsection{Mechanics of Nanocomposites with Agglomerated CNCs}

\begin{figure}[h]
    \centering
    \includegraphics[width=1.0\textwidth]{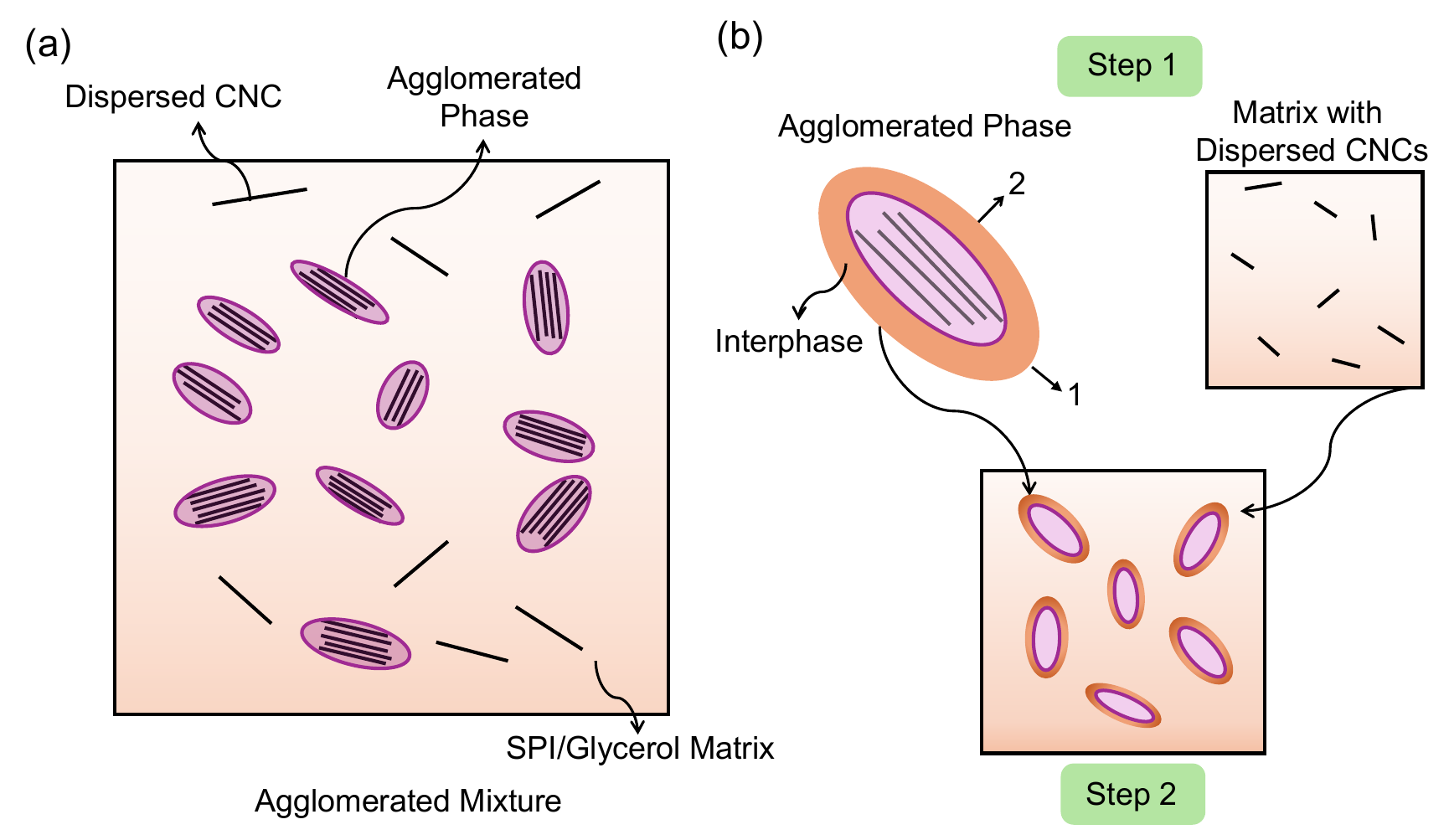}
    \caption{\textbf{Two-step micromechanical modeling approach for CNC-reinforced SPI-glycerol composites with agglomeration.} (a) Schematic of composite microstructure, showing CNCs partitioned into two states: agglomerated bundles and individually dispersed CNCs within the SPI/glycerol matrix. The interphase region between CNCs and matrix is highlighted, representing the zone of modified matrix properties due to CNC-matrix interactions. (b) The two-step homogenization process: Step 1 determined effective properties of the agglomerated CNC phase and the matrix containing dispersed CNCs separately. Step 2 combined these components using Mori-Tanaka theory to predict overall composite properties. This hierarchical approach enabled explicit consideration of both agglomeration effects and interphase contributions to composite stiffness.}
    \label{fig:3}
\end{figure}

TEM imaging reveals non-uniform dispersion of CNCs within the matrix: certain local regions contain ``agglomerated phases'' or CNC ``inclusions,'' while other regions show dispersed CNCs \cite{wang2024}. The CNC fibril bundles are randomly oriented, with an aspect ratio of approximately 2. To capture this, we used a two-step Mori-Tanaka model based on an RVE containing both agglomerated CNC bundles and dispersed CNCs within the SPI-glycerol matrix (\textbf{Fig. 3}). This hierarchical approach first determined the homogenized stiffness of the agglomerated phase and matrix separately, then calculated the homogenized stiffness of the complete nanocomposite using these effective properties. We used the two non-dimensional parameters of Shi, et al.\ \cite{shi2004}:
\begin{equation}
\xi = \frac{W_{\text{aggl}}}{W}, \quad \zeta = \frac{W_c^{\text{aggl}}}{W_c^{\text{aggl}} + W_c^{\text{disp}}}.
\end{equation}
where $W_c^{\text{aggl}}$ and $W_c^{\text{disp}}$ are the volumes of CNCs in the agglomerated phases and dispersed phases, respectively, $W_{\text{aggl}}$ the volume of the inclusions, and \(W\) is the total volume of the nanocomposite. Based on \textbf{Eqs. (4)} and \textbf{(5)}, the bulk and shear moduli of the agglomerated phase and medium are:
\begin{equation}
K_{\text{aggl}} = K_m + \frac{V_c \zeta (\delta_c - 3K_m \alpha_c)}{3(\xi - V_c \zeta + V_c \zeta \alpha_c)}, 
\end{equation}
\begin{equation}
G_{\text{aggl}} = G_m + \frac{V_c \zeta (\eta_c - 2G_m \beta_c)}{2(\xi - V_c \zeta + V_c \zeta \beta_c)}, 
\end{equation}
\begin{equation}
K_{\text{medium}} = K_m + \frac{V_c (\delta_c - 3K_m \alpha_c)(1 - \zeta)}{3[1 - \xi - V_c (1 - \zeta) + V_c (1 - \zeta) \alpha_c]}, 
\end{equation}
\begin{equation}
G_{\text{medium}} = G_m + \frac{V_c (1 - \zeta)(\eta_c - 2G_m \beta_c)}{2[1 - \xi - V_c (1 - \zeta) + V_c (1 - \zeta) \beta_c]}. 
\end{equation}

Finally, the elastic stiffness of the nanocomposite, considering  agglomeration effect, was estimated as:
\begin{equation}
K = K_{\text{medium}} + \frac{(K_{\text{aggl}} - K_{\text{medium}})\xi\phi}{(1 - \xi + \xi\phi)},
\end{equation}
\begin{equation}
G = G_{\text{medium}} + \frac{(G_{\text{aggl}} - G_{\text{medium}})\xi\phi}{(1 - \xi + \xi\phi)}.
\end{equation}

\subsection{Interphase Stiffening between CNCs and the Matrix}

A transitional interphase region between CNCs and matrix can enhance stress transfer and increase the effective matrix stiffness \cite{sharma2019}. While AFM indentation experiments have revealed gradients in elastic properties within these interphases \cite{pakzad2012,rahimi2017}, we adopted two simplifying assumptions: the interphase was treated as homogeneous and isotropic. These assumptions, while introducing some error, enabled analytical treatment of the problem.

For an interphase volume fraction relative to the agglomerated phase of $\rho_{\text{inter}} = {W_{\text{inter}}}/{W_{\text{aggl}}}$, where \( W_{\text{inter}} \) is the interphase volume, rule of mixtures (Voigt) estimates the effective bulk modulus \( K_{\text{medium}}^s \) and shear modulus \( G_{\text{medium}}^s \) of the medium were written:
\begin{equation}
K_{\text{medium}}^s = K_{\text{medium}}(1 - \xi - \rho_{\text{inter}}\xi) + K_{\text{inter}}\rho_{\text{inter}}\xi,
\end{equation}
\begin{equation}
G_{\text{medium}}^s = G_{\text{medium}}(1 - \xi - \rho_{\text{inter}}\xi) + G_{\text{inter}}\rho_{\text{inter}}\xi.
\end{equation}

The moduli of the matrix within the interphase increases with the degree of aggregation \cite{idumah2021, rahimi2017, sinko2015}, which was taken as proportional to the volume fraction of CNCs within the agglomerated phase ($W_c^{\text{aggl}}/{W}$ or $V_c \zeta)$.
To model this, we introduced a strengthening factor \( q_s \), representing the stiffening of the interphase due to strong CNC aggregation, so that the bulk modulus \( K_{\text{inter}} \) and shear modulus \( G_{\text{inter}} \) were written:
\begin{equation}
K_{\text{inter}} = q_s V_c \zeta K_m,
\end{equation}
\begin{equation}
G_{\text{inter}} = q_s V_c \zeta G_m.
\end{equation}
By substituting \( \xi(1 - \rho_{\text{inter}}) \) for \( \xi \) in \textbf{Eqs. (14)} and \textbf{(15)}, and  \( K_{\text{medium}}^s \) and \( G_{\text{medium}}^s \) for \( K_{\text{medium}} \) and \( G_{\text{medium}} \) in \textbf{Eqs. (16)} and \textbf{(17)}, we estimated the effects of both domain agglomeration and interphase stiffening on nanocomposite mechanics.

\section{Results and Discussion}

\begin{figure}[h]
\centering
\includegraphics[width=1.0\linewidth]{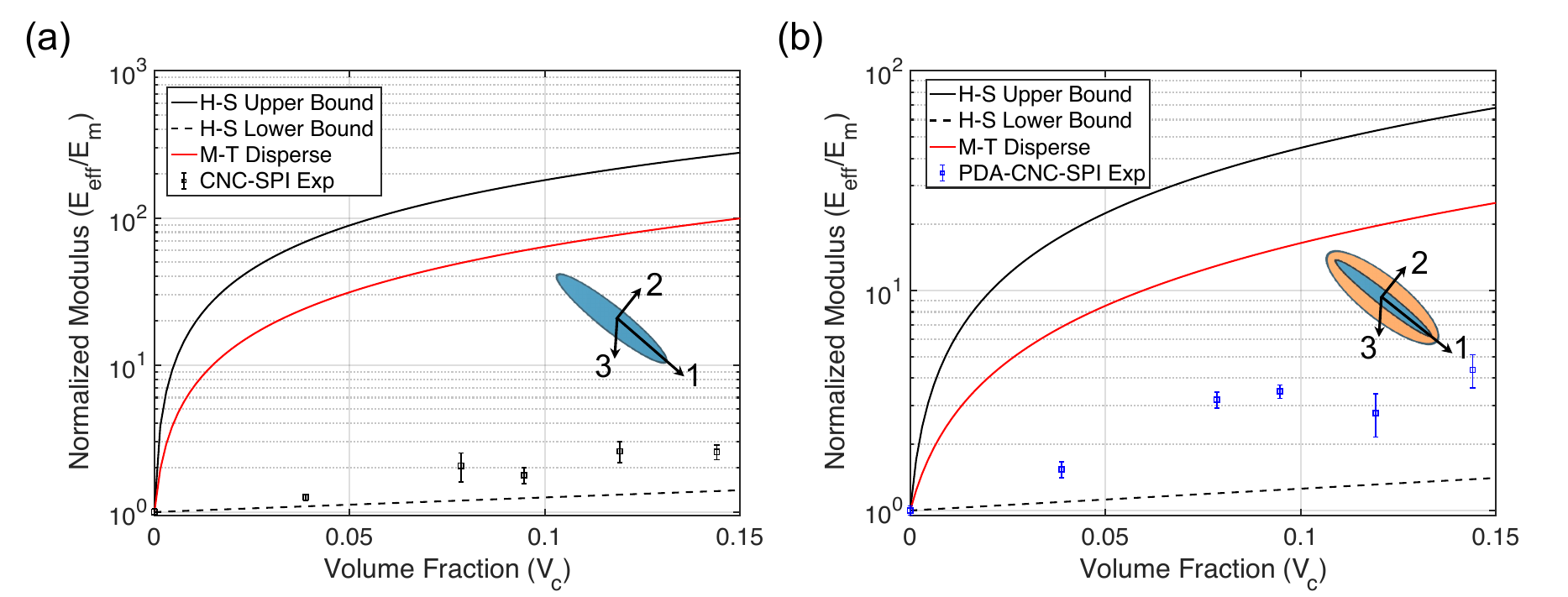}
\caption{\textbf{Comparison between experimental measurements and theoretical predictions for the elastic modulus of CNC-reinforced SPI-glycerol composites.} Results shown for (a) unmodified CNCs and (b) polyDOPA-modified CNCs. The Hashin-Shtrikman (H-S) bounds represent theoretical upper and lower limits for composite stiffness. The Mori-Tanaka (M-T) predictions assume perfect CNC dispersion. Experimental data (points) fall near the H-S lower bound, well below M-T predictions, suggesting the importance of microstructural features like CNC agglomeration and imperfect interfaces.}
\label{fig:4}
\end{figure}

\subsection{Comparison with Homogenization Bounds and Estimates}

We began by evaluating experimental measurements against theoretical bounds for composite stiffness. The Hashin-Shtrikman (H-S) bounds \cite{hashin1963} provide theoretical limits for the elastic properties of composites with dispersed isotropic, homogeneous inclusions (see complete formulation in \textbf{Appendix B}). 
Experimental data fell near the H-S lower bound (\textbf{Fig.~\ref{fig:4}}), with polyDOPA-modified CNCs yielding slightly higher stiffness despite the relatively compliant polyDOPA coating. 

This proximity to the lower bound is typical for particulate composites \cite{genin2009micromechanics}, but was unexpected for high aspect ratio fibers with high stiffness contrast between phases ($\sim10,000:1$). For well-dispersed, high-aspect-ratio inclusions, properties close to the H-S upper bound are expected \cite{weng1984,qiu1990}. Indeed, our Mori-Tanaka estimate for ideally dispersed CNCs (red solid line in \textbf{Fig.~\ref{fig:4}}) predicted stiffness an order of magnitude higher than that reported experimentally \cite{wang2024}. We explored two  potential causes: CNC agglomeration and imperfect interfaces. We therefore next examined how these features affect composite stiffness.

\begin{figure}[h]
\centering
\includegraphics[width=1.0\linewidth]{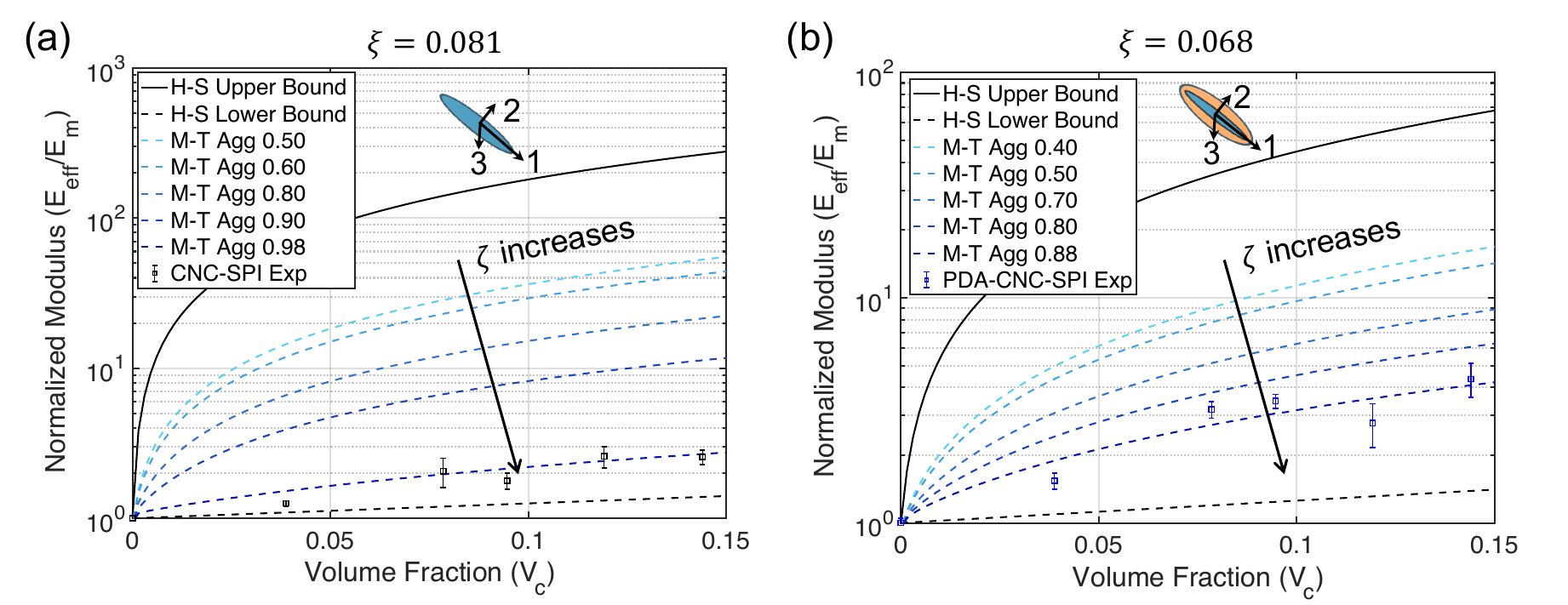}
   \caption{\textbf{Effect of CNC agglomeration on composite stiffness for (a)~unmodified CNCs and (b)~polyDOPA-modified CNCs.} Curves represent Mori-Tanaka (M-T) predictions for different values of the agglomeration parameter $\zeta$, which represents the fraction of CNCs in agglomerated versus dispersed states. Hashin-Shtrikman (H-S) bounds are shown for reference. Higher $\zeta$ values indicate increased agglomeration and correspond to lower composite stiffness. Experimental data points are best fitted with $\zeta = 0.98$ for unmodified CNCs and $\zeta = 0.88$ for polyDOPA-modified CNCs, with agglomeration parameters $\xi = 0.081$ and $0.068$ respectively. The lower $\zeta$ value for modified CNCs indicates improved dispersion, demonstrating the beneficial effect of polyDOPA surface treatment.}
   \label{fig:5}
\end{figure}

\subsection{Effects of CNC Agglomeration}

To study how domain agglomeration affects composite stiffness, we applied our model for several values of $\zeta$ (\textbf{Fig.~\ref{fig:5}}). 
TEM analysis \cite{wang2024} provided estimates of the agglomeration parameter $\xi$ (0.081 for unmodified CNCs and 0.068 for polyDOPA-CNCs). 
As expected, increasing $\zeta$ reduced composite stiffness. 
We then asked what values of $\zeta$ would be required to fit experimental data for the two CNC types. $\zeta = 0.98$ was required for unmodified CNCs, while $\zeta = 0.88$ was required for modified CNCs. 
This difference aligned with the expected effect of surface modification, with improved CNC-matrix affinity leading to better dispersion.

However, the required value of $\zeta = 0.98$ for unmodified CNCs exceeds experimental measurements ($\zeta_{\text{exp}} = 0.909 \pm 0.033$ at 10 wt\% CNCs) \cite{wang2024}. This discrepancy suggests our model may be compensating for other effects. Two possible explanations are: (1) CNC degradation during sonication, which can reduce crystallinity and introduce defects \cite{molnar2018, emenike2023}, and (2) matrix inhomogeneities, particularly variations in protein crystallinity that could locally reduce matrix stiffness. These effects would require artificially high values of $\zeta$ in the model to match experimental stiffness measurements.

\subsection{Role of Interphase Properties}

\begin{figure}[!]
\centering
\includegraphics[width=1.0\linewidth]{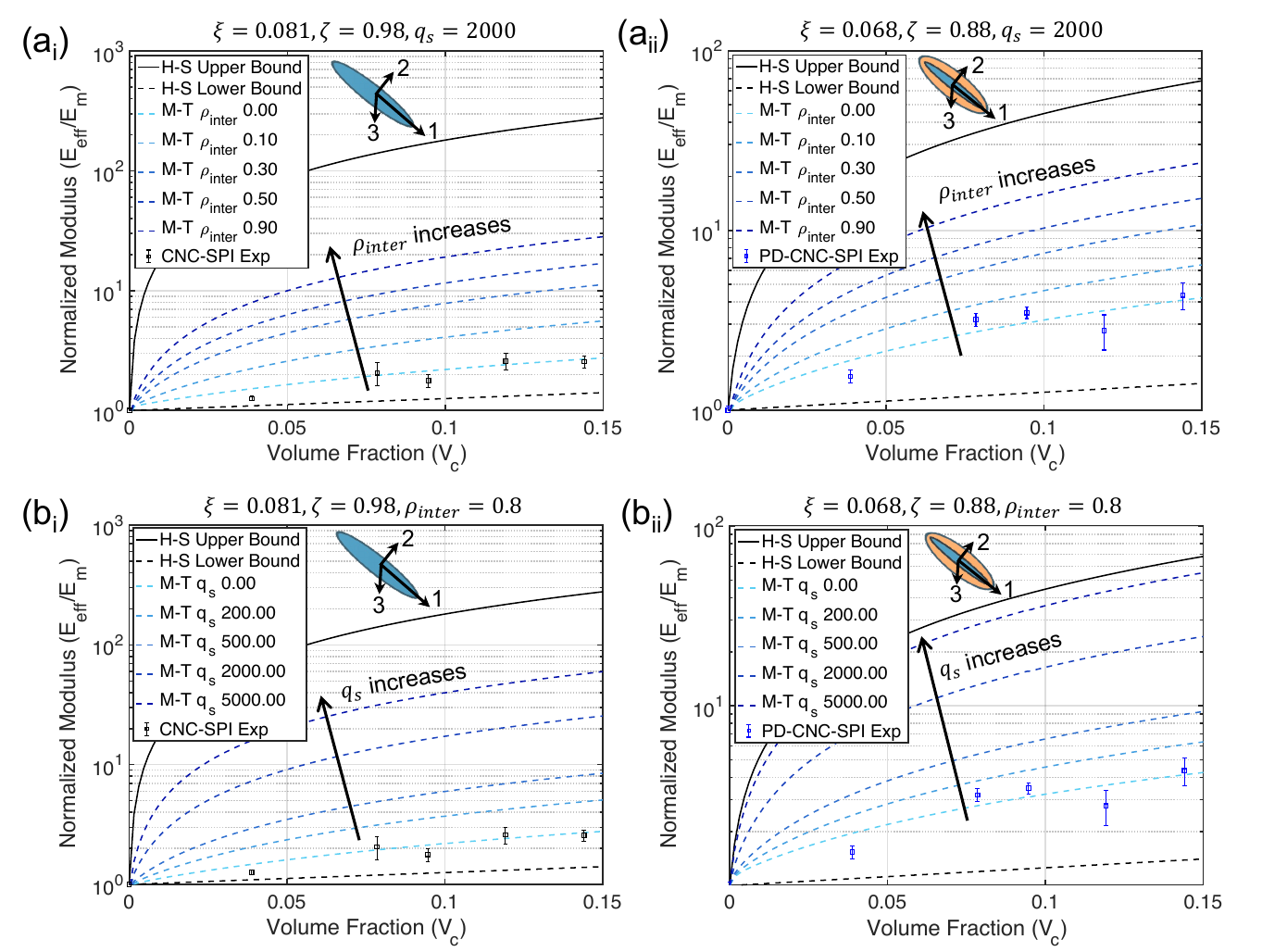}
   \caption{\textbf{Mori-Tanaka micromechanical modeling predictions incorporating both agglomeration and interphase effects.} (a) Impact of interphase volume ratio ($\rho_{\text{inter}}$) on composite stiffness for (a$_{\text{i}}$) unmodified CNCs and (a$_{\text{ii}}$) modified polyDOPA-CNCs, with fixed strengthening factor $q_{\text{s}} = 2000$. (b) Effect of interphase strengthening factor ($q_{\text{s}}$) on composite stiffness for (b$_{\text{i}}$) unmodified CNCs and (b$_{\text{ii}}$) modified polyDOPA-CNCs, with fixed interphase volume ratio $\rho_{\text{inter}} = 0.8$. Hashin-Shtrikman (H-S) bounds are included for reference. Agglomeration parameters are fixed at $\xi = 0.081$, $\zeta = 0.98$ for unmodified CNCs and $\xi = 0.068$, $\zeta = 0.88$ for polyDOPA-CNCs. Increasing either $\rho_{\text{inter}}$ or $q_{\text{s}}$ enhanced composite stiffness, with polyDOPA-modified CNCs showing greater sensitivity to these interphase parameters.}
\end{figure}

To study how interphase properties affect composite stiffness, we examined the effects of two key parameters: the interphase volume ratio $\rho_{\text{inter}}$ and the stiffening factor $q_s$. Varying $\rho_{\text{inter}}$ (\textbf{Fig. 6(ai)} and \textbf{(aii)}) revealed that larger interphase regions increased composite stiffness for both unmodified and modified CNCs. Similarly, increasing the stiffening factor $q_s$ at fixed $\rho_{\text{inter}}=0.8$ (\textbf{Fig. 6(bi)} and \textbf{(bii)}) enhanced composite stiffness, with polyDOPA-modified CNCs showing greater sensitivity to this effect.

These results suggested that surface modification works through two mechanisms: enhancing interfacial interactions and increasing the effective interphase volume. The combined effect can shift composite properties significantly closer to the H-S upper bound (\textbf{Fig. 6(bii)}). This finding has important implications for bio-nanocomposite design. Strategic modification of CNC surfaces could enable substantial improvements in mechanical performance without requiring changes to bulk material composition. The following section explores specific strategies for implementing these insights in bio-nanocomposites with better mechanical performance.

\subsection{Guidelines for Bio-nanocomposite Design}

\subsubsection{Matrix and CNC Stiffness Optimization}

The performance of bio-nanocomposites depends strongly on the stiffness contrast between the matrix (\(E_m\)) and CNCs (\(E_{\text{CNC}}\)). To understand this relationship quantitatively, we examined the ratio of predicted composite stiffness to the theoretical H-S upper bound, $E_{\text{M-T}}^{\text{aggl}}/E_{\text{H-S}}^{\text{Upper}}$ (\textbf{Fig. 7(ai, aii)}). This ratio provides a metric for how close a composite comes to achieving its theoretical maximum stiffness.

\begin{figure}[!]
\centering
\includegraphics[width=1.0\linewidth]{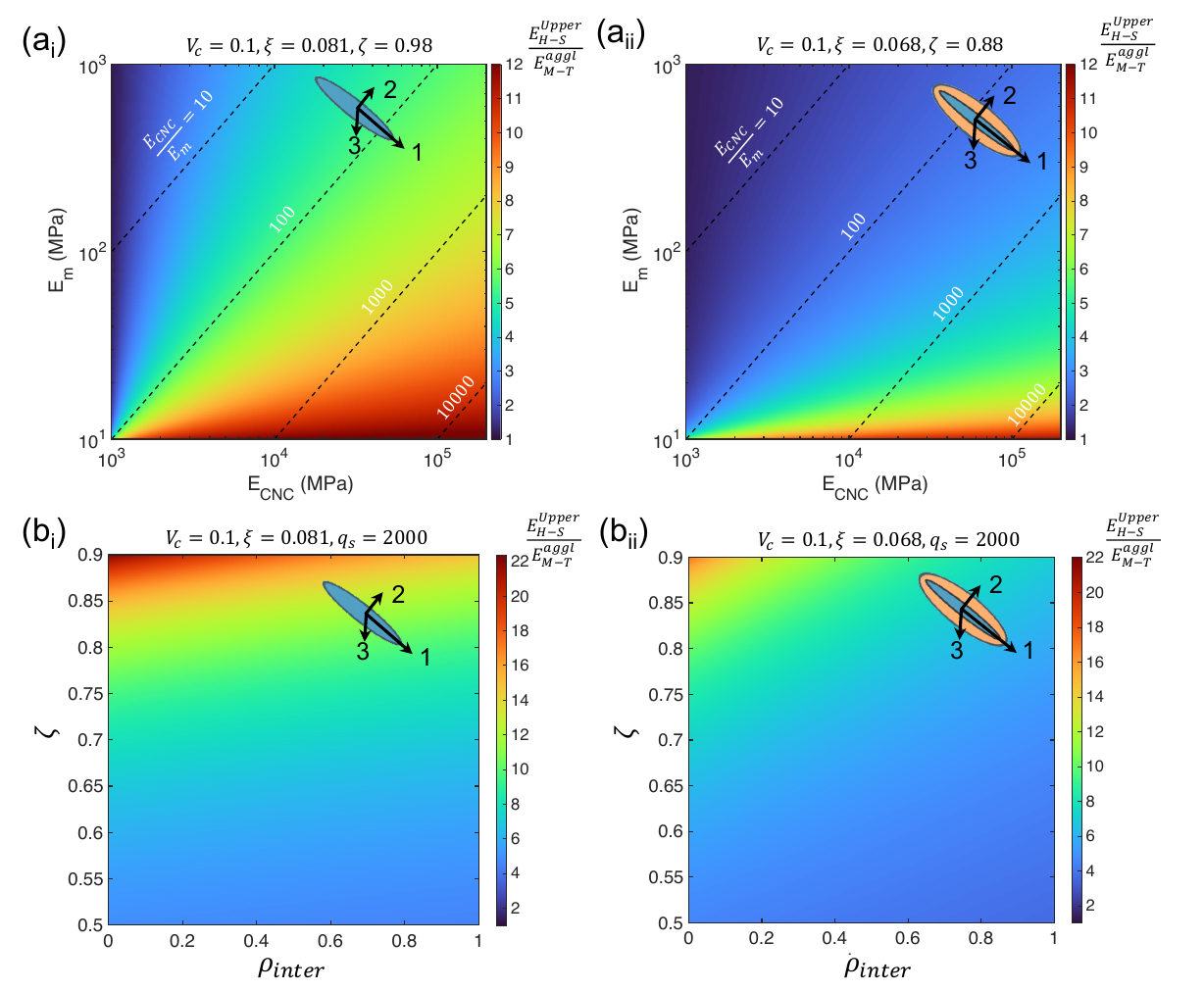}
    \caption{\textbf{Design maps for optimizing CNC-reinforced SPI nanocomposites.} (a) Colormaps showing the ratio of actual to theoretical maximum stiffness ($E^{\text{aggl}}_{\text{M-T}}/E^{\text{Upper}}_{\text{H-S}}$) as a function of matrix modulus ($E_{\text{m}}$) and CNC modulus ($E_{\text{CNC}}$) for (a$_{\text{i}}$) unmodified CNCs and (a$_{\text{ii}}$) modified CNCs. Contour lines indicate modulus ratios of $E_{\text{CNC}}/E_{\text{m}} = 10$, 100, 1000, and 10000. Parameters fixed at $V_{\text{c}} = 0.1$, with $\xi = 0.081$, $\zeta = 0.98$ for unmodified CNCs and $\xi = 0.068$, $\zeta = 0.88$ for modified CNCs. (b) Colormaps showing the influence of interphase volume ratio ($\rho_{\text{inter}}$) and agglomeration factor ($\zeta$) on normalized modulus for (b$_{\text{i}}$) unmodified CNCs ($\xi = 0.081$) and (b$_{\text{ii}}$) modified CNCs ($\xi = 0.068$), with $V_{\text{c}} = 0.1$ and $q_{\text{s}} = 2000$. These maps provide quantitative guidance for optimizing composite properties through control of matrix stiffness, CNC properties, and processing conditions.}
\end{figure}

Our analysis revealed three key findings. First, CNC reinforcement becomes more effective as matrix stiffness increases, with particularly dramatic improvements beyond $E_m = 100$ MPa. This effect is amplified in composites with polyDOPA-modified CNCs, suggesting surface modification enhances load transfer. Second, matrix stiffness ($E_m = 14.7$ MPa) in Wang et al. \cite{wang2024} falls well below optimal values, indicating substantial room for improvement through reduced water content, increased crosslinking, or decreased plasticizer concentration \cite{tian2018}.

Third, increasing CNC stiffness consistently improves composite performance, particularly in combination with stiffer matrices. This suggests two complementary paths to improvement: (1) optimizing acid hydrolysis conditions to reduce CNC defects, and (2) using high-temperature annealing to enhance CNC molecular alignment and eliminate surface defects \cite{matthews2011}. Based on these findings, we recommend targeting matrix stiffness above 100 MPa and CNC stiffness above 100 GPa to approach theoretical performance limits. These targets could be achieved through combined improvements in both matrix formulation and CNC processing.

\subsubsection{Strategies for Optimizing CNC Dispersion and Interfaces}
Phase maps examining the combined effects of agglomeration ($\zeta$) and interphase volume ($\rho_{\text{inter}}$) reveal two key pathways for improving composite performance (\textbf{Fig. 7(bi,bii)}). First, increased agglomeration ($\zeta$) consistently reduces load-bearing efficiency, highlighting the critical importance of achieving uniform CNC dispersion. While conventional mechanical stirring and sonication methods can achieve dispersion, they risk introducing CNC defects. Alternative approaches such as high-pressure homogenization \cite{lee2009} and surfactant-assisted processing \cite{shalauddin2022} may provide better dispersion while preserving CNC integrity.

Second, increasing the interphase volume ratio ($\rho_{\text{inter}}$) enhances composite stiffness, particularly when combined with good dispersion (low $\zeta$). This suggests that surface modifications promoting strong CNC-matrix interactions could significantly improve performance. Chemical approaches such as grafting and functionalization \cite{gomri2022} offer promising routes to optimize these interfacial regions. The synergistic relationship between dispersion and interface quality indicates that processing strategies should target both features simultaneously rather than addressing them independently.

\subsection{Model Limitations}
While our micromechanics framework captures key features of CNC-reinforced composites, several important simplifications merit attention in future work. Matrix heterogeneity remains unaddressed. The current model treats the SPI-glycerol matrix as homogeneous, neglecting variations in protein structure and crystallinity that affect both load transfer and interfacial interactions \cite{sui2021}. An iterative Mori-Tanaka approach incorporating these features could provide more accurate predictions, particularly at high CNC concentrations where matrix structure may be important.

Additionally, the parameters associated with  concentration dependence are difficult to characterize. $\xi$ was determined from TEM analysis at 10 Wt\% CNCs, but agglomeration behavior likely varies with concentration. Systematic TEM studies across multiple concentrations could establish this relationship, enabling more accurate modeling across the full range of practical CNC loadings.

Finally, the interface physics model here is a simplified representation, neglecting both property gradients and potential damage evolution. More sophisticated approaches incorporating multi-layer or gradient-based models could better capture interfacial complexity. Additionally, integrating cohesive zone models \cite{liu2024,guo2015} might enable prediction of progressive damage and failure, particularly important for understanding composite durability under cyclic loading.

\section{Conclusions}
This study provides insight into the mechanics of CNC-reinforced bio-nanocomposites through integrated experimental characterization and micromechanics modeling. Surface modification of CNCs with polyDOPA enhanced composite stiffness through two mechanisms: improved dispersion and stronger interfacial interactions. However, composite properties remained closer to theoretical lower bounds than expected for high-aspect-ratio reinforcements, pointing to opportunities for further optimization.

Our micromechanics framework revealed roles of both CNC aggregation and interphase properties. The model successfully captured experimental trends and provided quantitative targets for improvement: matrix stiffness above 100 MPa and CNC stiffness exceeding 100 GPa could enable order-of-magnitude improvements in composite performance. The analysis highlighted specific pathways for optimization: enhanced processing methods like high-pressure homogenization to achieve better CNC dispersion while minimizing defects, strategic surface modifications to control interface properties, and matrix modifications to increase stiffness through reduced plasticizer content or increased crosslinking.

These results inform rational design of sustainable bio-nanocomposites. While challenges remain, particularly in simultaneously optimizing multiple competing factors, this work demonstrates that strategic materials processing guided by mechanical modeling can enable substantial improvements in performance. Future work incorporating more sophisticated interface physics and concentration-dependent effects could further refine these design strategies, potentially enabling bio-based composites to rival conventional petroleum-based plastics.

\section*{Appendix A: CNC Transverse Isotropy has Negligible Effect on Composite Moduli}

\renewcommand{\thefigure}{A\arabic{figure}} 
\setcounter{figure}{0} 

\begin{figure}[H]
\centering
\includegraphics[width=0.6\linewidth]{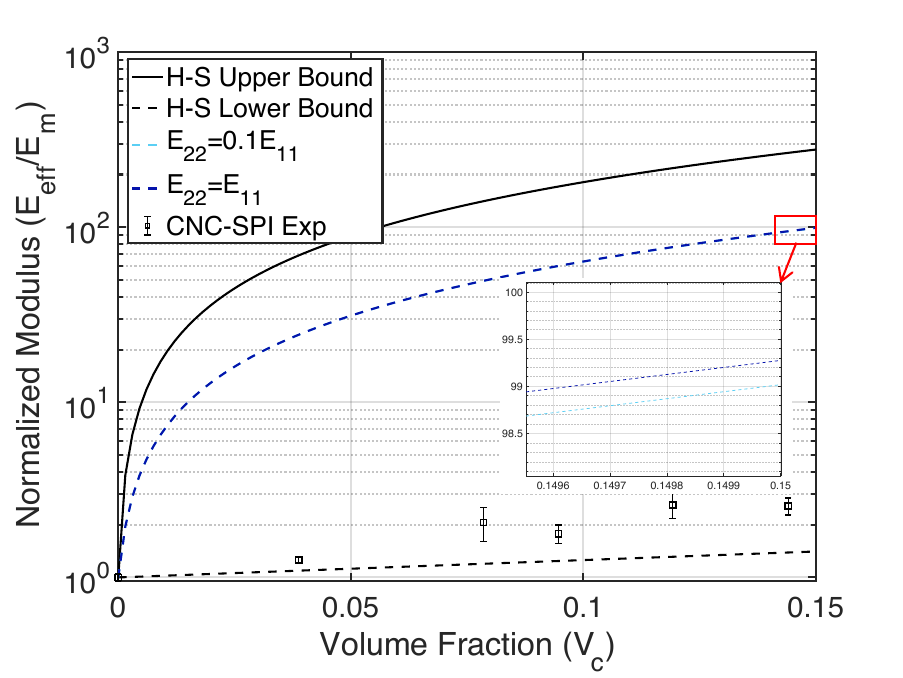}
\caption{\textbf{Effect of CNC transverse isotropy on the elastic modulus of CNC-reinforced SPI-glycerol composites.} Mori-Tanaka (M-T) predictions are shown for perfect CNC dispersion, considering two transverse isotropy cases: $E_{22}=0.1E_{11}$ and $E_{22}=E_{11}$. Results indicate that CNC transverse isotropy has a negligible effect on composite moduli.}
\label{fig:S1}
\end{figure}

\section*{Appendix B: Hashin-Shtrikman Bounds}

The bulk modulus and shear modulus from Hashin-Shtrikman lower bond (LB) and upper bond (UB) estimates are:
\begin{equation}
    K_{LB} = K_m + \frac{V_c}{\frac{1}{K_c - K_m} + V_m \left(K_m + \frac{4}{3}G_m \right)^{-1}}, \tag{S1}
\end{equation}
\begin{equation}
    K_{UB} = K_c + \frac{V_m}{\frac{1}{K_m - K_c} + V_c \left(K_c + \frac{4}{3}G_c \right)^{-1}}, \tag{S2}
\end{equation}
\begin{equation}
    G_{LB} = G_m + \frac{V_c}{\frac{1}{G_c - G_m} + \frac{6V_m(K_m + 2G_m)}{5G_m(3K_m + 4G_m)}}, \tag{S3}
\end{equation}
\begin{equation}
    G_{UB} = G_c + \frac{V_m}{\frac{1}{G_m - G_c} + \frac{6V_c(K_c + 2G_c)}{5G_c(3K_c + 4G_c)}}. \tag{S4}
\end{equation}
Here, $K_m$ and $K_c$ are the bulk moduli of SPI-glycerol matrix and CNCs, respectively;  $G_m$ and $G_c$ are the shear moduli of the matrix and CNCs, respectively;  and $V_m$ and $V_c$ are the volume ratios of the matrix and CNCs, respectively. Young’s modulus was be determined as 
$E = {9KG}/{(3K+G)}$.

\section*{Acknowledgements}
This work was funded by the NSF through grants OIA-2219142 and DMR 2105150.

\section*{Declaration of Interests}
The authors declare no competing interests.

\bibliographystyle{elsarticle-num-names}
\bibliography{references.bib}

\end{document}